\documentclass[%
 reprint,
superscriptaddress,
 amsmath,amssymb,
 prb,
]{revtex4-1}
\usepackage[dvipdfmx]{graphicx}
\usepackage{dcolumn}
\usepackage{bm}
\usepackage{color}

\begin{document}

\title{Time-dependent density functional theory for interaction  \\ 
of ultrashort light pulse with thin materials}

\author{Shunsuke Yamada}
\affiliation{Center for Computational Sciences, University of Tsukuba, Tsukuba 305-8577, Japan}
\author{Masashi Noda}
\affiliation{Center for Computational Sciences, University of Tsukuba, Tsukuba 305-8577, Japan}
\author{Katsuyuki Nobusada}
\affiliation{Institute for Molecular Science, Okazaki 444-8585, Japan}
\author{Kazuhiro Yabana}
\affiliation{Center for Computational Sciences, University of Tsukuba, Tsukuba 305-8577, Japan}

\date{\today}

\begin{abstract}
We present a comprehensive theoretical description for an irradiation of an ultrashort light pulse normally 
on thin materials based on first-principles time-dependent density functional theory. 
As the most elaborate scheme, we develop a microscopic description solving Maxwell equations for light electromagnetic fields 
and the time-dependent Kohn-Sham equation for electron dynamics simultaneously in  the time domain using a common spatial grid. 
We call it the microscopic Maxwell-TDDFT scheme.
We test this scheme for silicon thin films of various thickness, from a few atomic layers to a few tens of nm.
We show that the microscopic Maxwell-TDDFT scheme provides a satisfactory description incorporating 
electronic structure of thin films in the first-principles level, multiple reflections of the electromagnetic fields at the surfaces, 
and nonlinear light-matter interaction when the incident light pulse is strong.
However, the calculation becomes expensive as the thickness increases. 
We then consider two limiting cases of extremely thin and sufficiently thick films and develop approximate schemes. 
For the extremely thin case including two-dimensional atomic-layered materials, 
a two-dimensional macroscopic electromagnetism is developed:
a two-dimensional susceptibility is introduced for a weak field, while time evolution equation is derived for an intense field.
For sufficiently thick films, the microscopic Maxwell-TDDFT scheme is expected to coincide with a description utilizing ordinary 
macroscopic electromagnetism. We numerically confirm it comparing the calculated results:
For a weak field, a comparison is made with a description using the bulk dielectric susceptibility. 
For a strong field, a comparison is made with a multiscale Maxwell-TDDFT scheme which the authors'
group developed previously.
\end{abstract}
\maketitle

\section{Introduction}

Electronic and optical properties of thin materials have been extensively explored.
Thin materials of thickness comparable to or less than the wavelength of light 
are important for optical coating and devices\cite{Kats2016}. 
Quantum wells have been attracting interests for their unique optical properties 
due to the electron confinement in two-dimensional (2D) region\cite{Kelly1985}.
In the last decade, there have been an explosion of researches on 2D sheet materials 
such as graphene and transition metal 
dichalcogenide\cite{Novoselov2005,Nair2008,Britnell2013,Wang2012,Won2010}. 
In recent optical sciences, optical responses of thin materials using intense and ultrashort laser pulses are carried out
quite often to explore electron dynamics in the time domain\cite{Krausz2009,Bonaccorso2010,Calegari2016,Nicoletti2016}.

Under these experimental progresses, theories for the interaction of light with thin materials have been actively studied. 
Because electronic structures of thin materials are quite different from those of bulk materials,
quantum mechanical description for the electronic structure is indispensable.
For the light propagation in thin materials, theories utilizing response functions,
or equivalently, transfer matrix have been developed for weak pulses\cite{Saleh2007,Pendry1996,Katsidis2002,Zhan2013,Li2018,Gupta2018}.
However, for the interaction of intense and ultrashort light pulse, problems of
electronic structure and light propagation are no more separated.
Thus, it becomes important to develop a comprehensive theory that treats both electronic structures of 
thin materials and light propagation dynamics in thin materials simultaneously. 

This paper aims to report our progresses to develop a theoretical and computational
scheme that describes the interaction of ultrashort light pulse and thin materials, 
from one atomic layer to a few tens of nm thickness. 
In our theoretical description, we rely upon the time-dependent 
density functional theory (TDDFT) for electron dynamics\cite{Runge1984}.
TDDFT is an extension of the ordinary density functional theory that successfully describes ground states 
of electronic systems so that it describes electronic excitations and dynamics induced by an external field.
The basic equation that we solve is the time-dependent Kohn-Sham (TDKS) equation. 

Combined with a linear response theory, TDDFT has been quite successful to describe electronic excitations
and optical responses of molecules and solids\cite{Casida2009,Laurent2013,Paier2008,Yabana1996,Bertsch2000,Yabana2006}.
Solving the TDKS equation in real time, it can describe 
nonlinear and ultrafast electron dynamics induced by intense and ultrashort laser
pulses\cite{Petersilka1999,Calvayrac2000,Tong,Nobusada,Castro2004,Otobe2008,Shinohara,Miyamoto2010}.
We have furthermore developed a theoretical framework combining electron dynamics calculations with 
the macroscopic Maxwell equations that describe a propagation dynamics of light electromagnetic fields\cite{Yabana2012}.
It has been quite successful to describe extremely nonlinear optics such as  attosecond transient absorption
in solids\cite{Sommer2016,Lucchini2016}.

For optical responses of thin materials whose thickness is much less than the wavelength of the light pulse,
there is no doubt that description using ordinary macroscopic electromagnetism is no more useful.
We will develop a microscopic description, combining classical Maxwell equations that  describe the dynamics of 
light electromagnetic fields and TDKS equation that describes the electron dynamics. We use a common spatial
grid to solve the equations. As is anticipated, it is computationally expensive as the thickness of the thin
materials increase. We then explore approximate schemes. For responses to weak field, we will develop
macroscopic description, 2D description for extremely thin materials and ordinary macroscopic description
for sufficiently thick materials. For an intense field in which nonlinear light-matter interaction becomes
significant, we also develop macroscopic schemes for extremely thin and sufficiently thick cases.
To test the usefulness of our approach and to examine the validity of the approximates, 
we will apply our scheme to silicon (Si) thin films of thickness from a few atomic layers to a few tens of nm.

The construction of the paper is as follows.
In Sec.~\ref{sec:thory}, we present our formalism of microscopic and macroscopic descriptions for the coupled dynamics of electrons and electromagnetic fields. 
In Sec.~\ref{sec:silicon}, setup for calculations of Si thin films are explained. In Sec.~\ref{sec:low} and Sec.~\ref{sec:high}, calculated results are presented for low-intensity and high-intensity light pulses, respectively.
Finally, a conclusion is presented in Sec.~\ref{sec:conclusion}.

\section{Theoretical framework \label{sec:thory}}

\subsection{Set up of the problem}

\begin{figure}
    \includegraphics[keepaspectratio,width=\columnwidth]{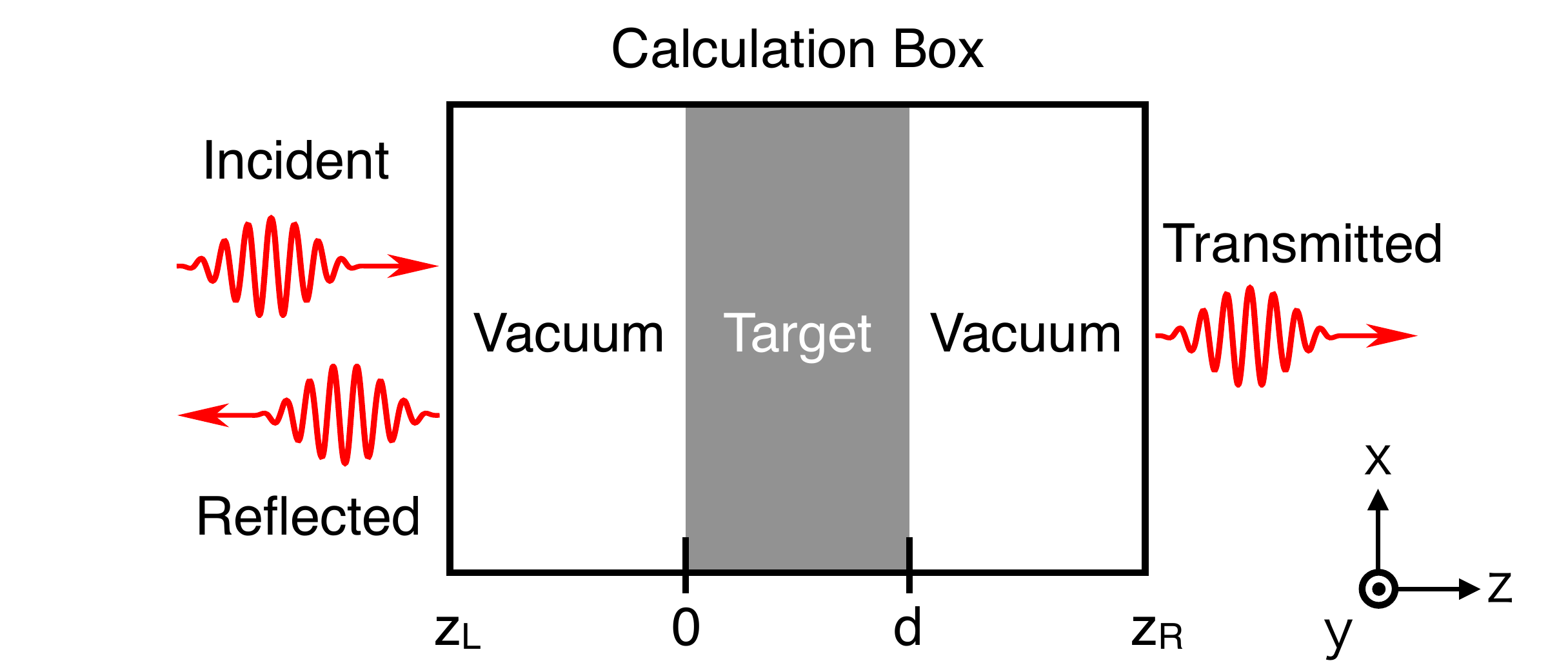}
    \caption{\label{fig:setup} 
    Setup of the problem. 
    The direction of the light propagation is along the $z$-axis. The thin material exists in the interval $z\in [0,d]$.
    $z_{\rm L}$ and $z_{\rm R}$ are the positions sufficiently far apart from the surface of the thin material such that the near field at these positions is negligible.
    }
\end{figure}

We consider an irradiation of a linearly polarized, ultrashort light pulse normally on a 
free-standing thin film of thickness $d$ in a vacuum.
We assume that the thin film is infinitely periodic in 2D and the thickness $d$ is much less than 
the wavelength of the incident light. 
We take a coordinate system as shown in Fig.~\ref{fig:setup}: The direction of the light propagation is taken along the
$z$-axis, the surfaces of the thin film are set at $z=0$ and $z=d$ planes, and the material exists in the interval $z\in [0,d]$.
The direction of the polarization of the incident light is taken along the $x$-axis.
We assume that the material of the thin film is isotropic in the 2D plane in macroscopic scale so that the direction of the 
macroscopic electric current induced by the light pulse is always parallel to the direction of the polarization 
of the light.

We describe the electromagnetic fields using a scalar and a vector potentials, $\phi({\bm r},t)$ and 
${\bm A}({\bm r},t)$, respectively. We assume that the near field of the thin film appears only in the spatial region
$z_{\rm L} < z < z_{\rm R}$ in Fig.~\ref{fig:setup}. In this region, the scalar and the vector potential show
complex spatial dependence, but have the same periodicity in $x$ and $y$ directions as that of the thin film.
We adopt the Gaussian unit system and utilize the Coulomb gauge throughout this paper. 

We assume that the ionization process from the surfaces of the thin film does not take place
so that the film is always neutral in charge. We also assume that there will be no dipole moment 
in the thin film in $z$ direction.
This is justified since the polarization direction of the incident pulse is chosen parallel to the surfaces of the film.
Then, the scalar potential exists in a spatial region of finite thickness from the thin film, $z_{\rm L} < z < z_{\rm R}$.

At the beginning of the calculation, the vector potential includes only the incident wave. It locates in $z<z_{\rm L}$
region and is described as the one-dimensional wave propagating in $z$ direction,
\begin{equation}
{\bm A}({\bm r},t) = \hat{\bm x} A^{\rm (i)} \left( t-\frac{z}{c} \right),
\label{eq:Ainitial}
\end{equation}
where $A^{\rm (i)}(t)$ specifies the time profile of the incident light pulse.

At a time long after the interaction, the vector potential consists of the reflected and the transmitted waves,
which exist in the spatial region $z<z_{\rm L}$ and $z>z_{\rm R}$, respectively. They are expressed as the one-dimensional waves
propagating along the $z$-axis as
\begin{equation}
{\bm A}({\bm r},t) = \hat{\bm x} A^{\rm (r)} \left(t+\frac{z}{c} \right) + \hat{\bm x} A^{\rm (t)} \left( t - \frac{z}{c} \right).
\label{eq:Afinal}
\end{equation}

In later sections, we will utilize the following quantity that shows the electromagnetic
energy passed through the $xy$-plane (specified by $z$) per unit area in the time interval from the infinite past to the time $t$,
\begin{equation}
e_{\rm EM}(z,t) = \int_{-\infty}^t dt'  \int_{\rm cell} \frac{dx dy}{\Omega} S_z(x,y,z,t'),
\label{eq:e_Poynting}
\end{equation}
where $S_z$ is the $z$ component of the Poynting vector and $\Omega$ is the area of the 2D unit cell.
Using this, we define the following quantities:
\begin{equation}
e^{\rm (r)}_{\rm EM}(t) = e^{\rm (i)}_{\rm EM}(z_{\rm L},t) - e_{\rm EM}(z_{\rm L},t),
\label{r-Poynting}
\end{equation}
\begin{equation}
e^{\rm (t)}_{\rm EM}(t) = e_{\rm EM}(z_{\rm R},t),
\label{t-Poynting}
\end{equation}
where $z_{\rm L}$ and $z_{\rm R}$ are the positions shown in Fig.~\ref{fig:setup}, 
and $e^{\rm (i)}_{\rm EM}(z,t)$ is identical to Eq.~(\ref{eq:e_Poynting}) but for the incident field $A^{\rm (i)}$.
$e^{\rm (r)}_{\rm EM}(t)$ is the energy of the reflected wave which passed through at position
$z_{\rm L}$ before the time $t$.
Similarly, $e^{\rm (t)}_{\rm EM}(t)$ is the energy of the transmitted wave that passed through at position $z_{\rm R}$ before the time $t$.
The energy per area of the incident wave at the infinite future, $e^{\rm (i)}_{\rm EM}(z,t \rightarrow + \infty)$, coincides with the fluence of the incident wave,
\begin{equation}
F^{\rm (i)}_{\rm EM} = e^{\rm (i)}_{\rm EM}(z,t \rightarrow \infty)= \frac{1}{4\pi c} \int_{-\infty}^{\infty} dt \left( \frac{dA^{\rm (i)}}{dt} \right)^2.
\label{eq:fluence}
\end{equation}

We define reflection, transmission, and absorption rates for the light pulse by $R=e^{\rm (r)}_{\rm EM}(t)/F^{\rm (i)}_{\rm EM}$, 
$T=e^{\rm (t)}_{\rm EM}(t)/F^{\rm (i)}_{\rm EM}$, and $A=1-R-T$, respectively.
Here we choose $t$ as the time just after the end of the pulse irradiation. 
In Sec.~\ref{sec:emission}, we will describe the $t$ dependence of the rates.

In the following subsections, we will develop theoretical methods that provide
$A^{\rm (r)}(t)$ and $A^{\rm (t)}(t)$ for a given $A^{\rm (i)}(t)$.
We will utilize the first-principles TDDFT to describe electron dynamics in thin films.
As the most comprehensive approach, we will first develop a scheme in which  the 
microscopic Maxwell equations for the electromagnetic fields and the TDKS equation for 
electron dynamics are simultaneously solved using a common spatial grid. 
As we will show, this approach is applicable to problems without imposing restrictions
on the intensity of the fields and on the thickness of the films.
However, it becomes computationally expensive as the thickness increases.

We will then develop approximate macroscopic descriptions introducing a coarse-graining.
We classify the problem into four types according to the strength of the electromagnetic fields, 
linear and nonlinear regimes, and according to the thickness of the materials, extremely thin 
and sufficiently thick cases.

\subsection{Microscopic Maxwell-TDDFT formalism \label{sec:micro}}

As a comprehensive and elaborate method, we develop the microscopic Maxwell-TDDFT scheme
in this subsection. In this scheme, we solve the microscopic Maxwell equations for the electromagnetic 
fields and the TDKS equation for the electron dynamics simultaneously, using a common spatial grid. 

We describe the propagation of the electromagnetic fields using the microscopic Maxwell equations.
The vector potential ${\bm A}({\bm r},t)$ and the scalar potential $\phi({\bm r},t)$ satisfy the following equation,
\begin{equation}
    \left( \frac{1}{c^2} \frac{\partial^2}{\partial t^2} - \nabla^2 \right) {\bm A}({\bm r},t) 
    + \frac{1}{c} \frac{\partial}{\partial t} \nabla \phi({\bm r},t) =  \frac{4\pi}{c} {\bm j}({\bm r},t),
\label{eq:Aeq}
\end{equation}
where ${\bm j}({\bm r},t)$ is the electric current density.
The scalar potential also satisfies the Poisson equation,
\begin{equation}
    \nabla^2 \phi({\bm r},t) = - 4\pi \rho ({\bm r},t),
    \label{eq:poisson}
\end{equation}
where $\rho({\bm r},t)$ is the charge density. 
The charge density is composed of ionic and electronic contributions. We will freeze the position of ions and 
there appears no electric current density coming from ionic motion.
In our problem described in the previous subsection, the charge density and the electric current 
density have the same 2D periodicity as that of the thin film. 
The scalar and the vector potentials 
have the same periodicity as well. 

Since the whole system has the 2D periodicity at each time, 
electron dynamics in the thin film can be described using Bloch orbitals $\{ u_{n{\bm k}}({\bm r},t) \}$, which are periodic in the $xy$-plane. 
${\bm k}=(k_x, k_y, 0)$ represents the 2D crystalline momentum vector.
Before the incident field arrives, the electronic state is given by the ground state solution of the static
DFT calculation.

The evolution of the Bloch function $u_{n{\bm k}}({\bm r},t)$ is governed by the following TDKS equation:
\begin{eqnarray}
    i\hbar \frac{\partial u_{n{\bm k}}({\bm r},t)}{\partial t}
    &=& \biggl\{ \frac{1}{2} \left( -i\hbar \nabla + \hbar{\bm k} + \frac{e}{c} {\bm A}({\bm r},t) \right)^2 
    - e\phi({\bm r},t)  \nonumber \\ 
    && + \delta V_{\rm{ion}}({\bm r}) + V_{\rm{xc}}({\bm r},t) \biggr\} u_{n{\bm k}}({\bm r},t),
    \label{eq:TDKS}
\end{eqnarray}
where $V_{\rm{xc}}({\bm r},t)$ is the exchange-correlation potential.
We use a norm-conserving pseudopotential for the electron-ion potential \cite{Troullier1991}.
$\delta V_{\rm ion}({\bm r})$ is the nonlocal part of the pseudopotential. 
The method of Ref.~\onlinecite{Pickard2003} is used to incorporate the gauge dependence for the 
nonlocal term of the pseudopotential \cite{Varsano2009,Ismail-Beigi2001,Pickard2003}.
The local part of the electron-ion potential is included in the scalar potential $\phi({\bm r},t)$.

The electron matter current density ${\bf j}_{\rm e}({\bf r},t)$ and the electron density distribution $n_{\rm e}({\bf r},t)$ 
are obtained from the Bloch orbitals as follows:
\begin{eqnarray}
    n_{\rm e}({\bm r},t) &=& \sum_{n{\bm k}}^{\rm{occ}} |u_{n{\bm k}}({\bm r},t)|^2 , 
    \label{eq:density} \\
    {\bm j}_{\rm e}({\bm r},t) &=& {\rm Re} \sum_{n{\bm k}}^{\rm occ} u_{n{\bm k}}^{\ast}({\bm r},t)
    \Bigl( -i\hbar \nabla \nonumber \\
    && + \hbar{\bm k} + \frac{e}{c} {\bm A}({\bm r},t) \Bigr) u_{n{\bm k}}^{}({\bm r},t),
    \label{eq:current}
\end{eqnarray}
where the indices run over the occupied bands in the ground state.
Since the nonlocal pseudopotential does not commute with the coordinate operator, it contributes to the electric 
current \cite{Bertsch2000}.
In Eq.~(\ref{eq:current}), we ignore the contribution from the nonlocal part of $\delta V_{\rm ion}$.
The electric current density and the charge density are given by ${\bf j}({\bf r},t)=-e{\bf j}_{\rm e}({\bf r},t)$ and $\rho({\bf r},t) = \rho_{\rm ion}({\bf r}) - e n_{\rm e}({\bf r},t)$,  respectively.
Here $\rho_{\rm ion}({\bf r})$ is the charge density of the ion cores.

Solving Eqs.~(\ref{eq:Aeq})--(\ref{eq:current}) simultaneously, we obtain the solution:
time evolution of the light electromagnetic fields and the electronic excitations inside the thin film.
This microscopic Maxwell-TDDFT scheme is quite comprehensive and satisfactory, having 
a number of favorable features: 
The electronic structure of the thin film is included in the first-principles level.
The multiple reflections of light electromagnetic fields at the surfaces of the thin film are treated adequately.
We may use any time profile for the incident light pulse. 
Since the TDKS equation is solved without any perturbative approximation, 
the scheme can treat nonlinear interaction that should be significant for a strong incident light.
The scheme can be applicable to thin films of various thickness, from single atomic-layer 
materials to thick films, although the computational cost increases as the thickness increases.

\subsection{2D macroscopic description: thin limit \label{sec:2D_macro}}

In this subsection, we develop an approximation to the microscopic Maxwell-TDDFT scheme
introducing a 2D coarse graining. This is appropriate for an extremely thin material, and 
we call it the 2D macroscopic description.

We introduce the 2D coarse graining as follows:
We assume a zero-thickness limit for the thin material and make a coarse graining of 
the Maxwell equation for the vector potential, Eq.~(\ref{eq:Aeq}), 
while the Maxwell equation for the scalar potential, Eq.~(\ref{eq:poisson}), is left as microscopic. 
The electric current density in Eq.~(\ref{eq:Aeq}) is approximated as
\begin{equation}
{\bm j}({\bm r},t) \simeq \hat{\bm x}\delta(z) \tilde J(t).
\label{eq:jmacro}
\end{equation}
The 2D macroscopic current density $\tilde J(t)$ is related to the microscopic one by
\begin{equation}
\tilde J(t) =  \int_{\rm cell} \frac{dx dy}{\Omega} \int dz {\bm j}({\bm r},t).
\label{eq:jav}
\end{equation}
The scalar potential term in Eq.~(\ref{eq:Aeq}) vanishes
after averaging over the unit cell area because of the periodicity. 
Under the above assumption and approximation, Eq.~(\ref{eq:Aeq}) reduces to the one-dimensional 
macroscopic Maxwell equation,
\begin{equation}
\frac{1}{c^2} \frac{\partial^2}{\partial t^2} A(z,t) - \frac{\partial^2}{\partial z^2} A(z,t)
= \frac{4\pi}{c} \delta(z) \tilde J(t).
\label{eq:A2eqD}
\end{equation}

In the TDKS equation (\ref{eq:TDKS}), we replace the vector potential, ${\bm A}({\bm r},t)$ with 
$\hat{\bm x} A(z=0,t)$,
\begin{eqnarray}
    i\hbar \frac{\partial u_{n{\bm k}}({\bm r},t)}{\partial t}
    &=& \biggl\{ \frac{1}{2} \left( -i\hbar \nabla + \hbar{\bm k} + \frac{e}{c} \hat{\bm x}A(0,t) \right)^2 
    - e\phi({\bm r},t)  \nonumber \\ 
    && + \delta V_{\rm{ion}}({\bm r}) + V_{\rm{xc}}({\bm r},t) \biggr\} u_{n{\bm k}}({\bm r},t),
    \label{eq:TDKS2D}
\end{eqnarray}

Equations~(\ref{eq:jav})-(\ref{eq:TDKS2D}) constitute our 2D macroscopic description.
Equation (\ref{eq:A2eqD}) includes delta function in the right hand side.
We treat it as a boundary value problem among $A^{\rm (i)}(t)$, $A^{\rm (r)}(t)$, and $A^{\rm (t)}(t)$, as below.
At $z=0$, $A(z,t)$ is continuous and its derivative satisfies
\begin{equation}
- \left. \frac{\partial A}{\partial z} \right\vert_{z=0_+} + \left. \frac{\partial A}{\partial z} \right\vert_{z=0_-} 
= \frac{4\pi}{c} \tilde J(t).
\label{eq:2Dderivative}
\end{equation}
The continuity condition gives us
\begin{equation}
A^{\rm (i)}(t) + A^{\rm (r)}(t) = A^{\rm (t)}(t).
\label{eq:2Dcont}
\end{equation}
Eqs.~(\ref{eq:2Dderivative})-(\ref{eq:2Dcont}) bring us the equation for $A^{\rm (t)}(t)$,
\begin{equation}
\frac{dA^{\rm (t)}}{dt} = \frac{dA^{\rm (i)}}{dt} + 2\pi \tilde J[A^{\rm (t)}](t),
\label{eq:2DAmacro}
\end{equation}
where we denote $\tilde J(t)$ as $\tilde J[A^{\rm (t)}](t)$ to stress that $\tilde J(t)$ is
determined by the vector potential at $z=0$ that is equal to $A^{\rm (t)}(t)$.
$A^{\rm (r)}(t)$ is obtained by $A^{\rm (r)}(t) = A^{\rm (t)}(t) - A^{\rm (i)}(t)$.

Up to this stage, we do not make any assumptions on the strength of the electromagnetic fields.
When the incident light pulse is strong, we need to solve Eq.~(\ref{eq:2DAmacro}) as an initial value problem.
For a given incident pulse $A^{\rm (i)}(t)$, we obtain $A^{\rm (t)}(t)$ by solving Eq.~(\ref{eq:2DAmacro}) with the initial condition,
$A^{\rm (t)}(t=-\infty) = 0$. 

When the incident light is sufficiently weak, we may make use of a linear response treatment.
In the linear regime, TDKS equation (\ref{eq:TDKS2D}) provides the following constitutive relation,
\begin{equation}
\tilde J(t) = \int^t dt' \tilde \sigma(t-t') E(t') = -\frac{1}{c} \int^t dt'  \tilde \sigma(t-t') \frac{dA(0,t')}{dt'},
\label{eq:J_sigmaE}
\end{equation}
where $E(t)$ is the electric field at the thin material and $\tilde \sigma$ is the 2D conductivity of the material.
Taking Fourier transformation of Eqs.~(\ref{eq:2DAmacro}) and (\ref{eq:J_sigmaE}), we obtain
\begin{equation}
\frac{A^{\rm (t)}(\omega)}{A^{\rm (i)}(\omega)} = \left(1+\frac{2\pi}{c} \tilde \sigma(\omega) \right)^{-1},
\label{eq:trans_coef}
\end{equation}
where $\tilde \sigma(\omega)$ is defined as the Fourier transformation of $\tilde \sigma(t)$.
Thus, for a weak field, it is sufficient to calculate the 2D conductivity $\tilde \sigma(\omega)$
to get the relation between $A^{\rm (r)}(t)$, $A^{\rm (t)}(t)$, and $A^{\rm (i)}(t)$.
We note that the definition of the 2D conductivity is consistent with that in literature\cite{Nair2008,Mak2008,Stauber2008,Zhan2013,Matthes2014}, and the reflection coefficient for the normal incidence derived from Eq.~(\ref{eq:trans_coef}) 
agrees with that in previous studies\cite{Sernelius2012,Matthes2014,Li2018}.

The 2D conductivity, $\tilde \sigma(\omega)$ can be efficiently calculated by solving the TDKS equation
 (13) in real time, extending the method developed in Ref.~\onlinecite{Bertsch2000,Yabana2012}.
As the vector potential which distorts the electronic system instantaneously, we use the following form,
\begin{equation}
A(t) = \hat {\bm x} A_0 \theta(t),
\end{equation}
where $\theta(t)$ is the step function. 
This amounts to shift the ${\bm k}$ vector by a small amount at $t=0$.
This instantaneous distortion causes a macroscopic 2D electric current density, $\hat{\bm x} \tilde J(t)$.
The conductivity is obtained by the time-frequency Fourier transformation of $\tilde J(t)$,
\begin{equation}
\tilde \sigma(\omega) = -\frac{c}{A_0} \int dt e^{i\omega t} \tilde J(t).
\label{eq:2Dsigma}
\end{equation}

Also, the 2D electric susceptibility $\tilde \chi(t)$ is defined by
\begin{equation}
\tilde P (t) = \int^t dt' \tilde \chi(t-t') E(t'),
\end{equation}
where $\tilde P (t)=\int^t dt' \tilde J(t')$ is the 2D dielectric polarization density.
The Fourier transformation of $\tilde \chi(t)$ is related to the 2D conductivity by
\begin{equation}
\tilde \chi (\omega) = \frac{i}{\omega}\tilde \sigma(\omega).
\label{eq:2D_chi}
\end{equation}
We note that the definition of $\tilde \chi (\omega)$ is consistent with the polarizability per unit area of graphene in previous studies \cite{Yang2009,Iida2018}. 

\subsection{3D macroscopic description: thick limit \label{sec:3D_macro}}

When the thin material is sufficiently thick, it will be reasonable to use an ordinary macroscopic
electromagnetism. For a thin film that consists of a material characterized by the complex index 
of  refraction, $N(\omega)=\sqrt{1+4\pi \chi(\omega)}$, the reflection, transmission, and 
absorption rates are given by\cite{Dressel2002}
\begin{equation}
R=\left| \frac{(1-N^2)(1-e^{-2i\omega N d /c})}{(1-N)^2- (1+N)^2 e^{-2i\omega N d /c}} \right|^2,
\label{eq:3DmacroR}
\end{equation}
\begin{equation}
T=\left| \frac{ - 4 N e^{-i\omega N d /c} }{(1-N)^2- (1+N)^2 e^{-2i\omega N d /c}} \right|^2,
\label{eq:3DmacroT}
\end{equation}
\begin{equation}
A=1-R-T,
\label{eq:3DmacroA}
\end{equation}
respectively.
Calculating the index of refraction (electric susceptibility) using TDDFT, we may regard this calculation
as an approximation to our microscopic Maxwell-TDDFT scheme when the incident pulse is weak. 

However, we should note that Eqs.~(\ref{eq:3DmacroR})-(\ref{eq:3DmacroA}) will never coincide exactly with the results of the
microscopic Maxwell-TDDFT scheme.
In the ordinary macroscopic electromagnetism, we assume a local constitutive relation, 
$P({\bm r},t) = \int dt' \chi(t-t') E({\bm r},t')$. 
On the other hand, our microscopic Maxwell-TDDFT scheme in the linear regime amounts to 
assuming a constitutive relation that is nonlocal in space\cite{Cho1991,Cho2003},
$P({\bm r},t) = \int dt' \int d{\bm r'} \chi({\bm r}, {\bm r'}, t-t') E({\bm r'},t')$,
since Bloch orbitals are assumed to be extended over the whole thin film.
In other words, our microscopic Maxwell-TDDFT scheme incorporates nonlocal optical response 
that is usually ignored in the macroscopic electromagnetism.

When the incident electric field is strong, we can no more describe the propagation of
the light electromagnetic fields using the electric susceptibility. We can instead use the multiscale 
Maxwell-TDDFT scheme that we developed previously \cite{Yabana2012}. 
Since we will later show a comparison 
between microscopic and multiscale Maxwell-TDDFT schemes, we briefly explain the formalism.

We describe a propagation of a light pulse using the macroscopic Maxwell equation.
For the one-dimensional propagation along $Z$ coordinate, we use
\begin{equation}
\frac{1}{c^2} \frac{\partial^2}{\partial t^2} A_Z(t) - \frac{\partial^2}{\partial Z^2} A_Z(t)
=  \frac{4\pi}{c} J_Z(t).
\label{eq:Aeqmultiscale}
\end{equation}
We assume that the electric current density at the position $Z$, $J_Z(t)$, is determined by the Bloch orbitals 
at the position $Z$, $u_{n{\bm k},Z}({\bm r},t)$, which is taken to be periodic in three dimension. 
It satisfies the TDKS equation,
\begin{eqnarray}
    &&i\hbar \frac{\partial u_{n{\bm k},Z}({\bm r},t)}{\partial t}
    = \biggl\{ \frac{1}{2} \left( -i\hbar \nabla + \hbar{\bm k} + \frac{e}{c} \hat{\bm x}A_Z(t) \right)^2 \nonumber \\
    &&- e\phi_Z({\bm r},t)   
     + \delta V_{\rm{ion}}({\bm r}) + V_{\rm{xc}}({\bm r},t) \biggr\} u_{n{\bm k},Z}({\bm r},t).
    \label{eq:TDKS-3D}
\end{eqnarray}
The electron dynamics calculation is carried out simultaneously for each $Z$ point in this multiscale scheme. 

\subsection{Numerical implementation}

We implement the method in the open-source software SALMON (Scalable Ab-initio Light-Matter simulator for
Optics and Nanoscience) \cite{SALMON2018} which has been developed in our group and is available
from the website, \cite{SALMON_web}.

We describe both the Bloch orbitals and the electromagnetic potentials utilizing a real-space finite-difference
scheme on a common uniform Cartesian grid in three dimensions.
To describe the electronic structure, we utilize the three-dimensional Bloch orbitals containing the thin material in the slab approximation
in which sufficiently thick vacuum layers are prepared in both sides of the thin material. 
For the vector potentials ${\bm A}({\bm r},t)$, 
periodic boundary conditions are imposed on the $xy$-plane and the Mur absorbing boundary condition \cite{Mur1981} 
is adopted in $z$ direction. 

Before we start the time evolution calculation, we carry out the ground state calculation of the thin material in the static
density functional theory that will be used as the initial Bloch orbitals.
In the time evolution calculation of the Bloch orbitals, we use a fourth-order expansion of the TDKS evolution 
operator \cite{Yabana1996,Yabana2006}:
\begin{equation}
    u_{n{\bm k}}({\bm r},t+\Delta t_{\rm e}) = \sum^4_{N=0} \frac{(-iH(t)\Delta t_{\rm e})^N}{N!} u_{n{\bm k}}({\bm r},t),
\end{equation} 
where $H(t)$ is the Kohn-Sham Hamiltonian of Eq.~(\ref{eq:TDKS}) at the time $t$ and $\Delta t_{\rm e}$ is the time step. 
The spatial differentiation operator is approximated by the fourth-order finite difference. 
For the time evolution of the vector potential, we use the second-order finite difference formula, 
\begin{eqnarray}
    {\bm A}({\bm r},t+\Delta t_{\rm EM})&=& 2{\bm A}({\bm r},t) - {\bm A}({\bm r},t- \Delta t_{\rm EM}) \nonumber \\
    &+& c^2 (\Delta t_{\rm EM})^2 \biggl( [\nabla^2 {\bm A}]({\bm r},t) 
    \nonumber \\
    &-& \frac{1}{c} \left[ \frac{\partial \nabla \phi}{\partial t }  \right]({\bm r},t) + \frac{4\pi}{c}{\bm j}({\bm r},t) \biggr),
    \label{eq:maxwell_step}
\end{eqnarray}
where $\Delta t_{\rm EM}$ is the time step for the Maxwell equation. 
The spatial differentiation is approximated by the second-order finite difference for the electromagnetic potentials.
Besides, the Poisson equation [Eq.~(\ref{eq:poisson})] is solved at each time step by using the ordinary discrete Fourier transform method.

In the time evolution calculation, we should be aware that the typical speed of the propagation is very different
between the electromagnetic fields and the electrons. This can be understood by the value of the speed of light, 
$c \approx 137$ in atomic units. In the explicit schemes we adopted for the time evolution calculations, 
we need to use quite different values for the time steps, $\Delta t_{\rm e}$ and $\Delta t_{\rm EM}$, to make the calculations stable. 
In later calculation, we will adopt $\Delta t_{\rm e} = 100 \Delta t_{\rm EM}$.
In solving Eq.~(\ref{eq:maxwell_step}), we update ${\bm A}$ using the time step of $\Delta t_{\rm EM}$, while ${\bf j}({\bf r},t)$ and 
$[\partial \nabla \phi/\partial t]({\bf r},t)$ are updated in the time step of $\Delta t_{\rm e}$. We use the same values of
${\bf j}({\bf r},t)$ and $[\partial \nabla \phi/\partial t]({\bf r},t)$ in updating ${\bm A}$ until the electronic 
system is updated.  
The scalar potential $\phi({\bf r},t)$ is also updated by solving Eq.~(\ref{eq:poisson})
when the electronic system is updated. 

In the 2D macroscopic description, calculation of the Bloch orbitals is achieved in the same way as that described above.
The time evolution calculation of the vector potential, Eq.~(\ref{eq:2DAmacro}), is achieved using a simple formula,
\begin{eqnarray}
&&A^{\rm (t)}(t+\Delta t_{\rm EM}) = A^{\rm (t)}(t-\Delta t_{\rm EM}) \nonumber\\
&+& 2 \Delta t_{\rm EM} \left( \frac{dA^{\rm (i)}(t)}{dt} + 2 \pi \tilde J[A^{\rm (t)}](t) \right).
\end{eqnarray}

\section{Silicon nano film \label{sec:silicon}}

To test our formalism, we apply our method for a thin film of silicon (Si) with (001) surfaces.
In this section, we explain the set up of the calculation and show a typical calculation.

We make calculations for films of thickness $d=na$ with $n=1$, $3$, $10$, and $50$, 
where $a=5.43$ {\AA} is the lattice constant of Si in the cubic unit cell containing eight atoms.
Thus the thickness of the films are from $d=a \sim 0.54$ nm to $d=50a \sim 27$ nm.
The atomic positions are not relaxed but are set to those of the bulk silicon.
The dangling bonds that appear surfaces are terminated with hydrogen atoms. 

We set the lengths of the sides of the computational box area as the lattice constant,
$L_{x,y}=a$ in $x$- and $y$-directions and $L_z=d+4a$ for $z$-direction. 
Namely, we attach the vacuum layers of thickness $2a$ to both sides of the Si film, $z_{\rm L}=-2a$ and $z_{\rm R}=d+2a$ in Fig.~\ref{fig:setup}. 
At the two boundaries, $z=z_{\rm L}$ and $z_{\rm R}$, we impose the periodic boundary 
conditions for the Bloch orbitals $u_{n{\bm k}}$ and the scalar potential $\phi$,  and the absorbing boundary condition for the vector potential ${\bm A}$.
Here the right-going incident wave $A^{\rm (i)}$ is incorporated in the absorbing boundary condition at $z=z_{\rm L}$ as the external field, and the left-going reflected wave is captured by this boundary.
The grid spacing for the spatial finite-difference calculation is chosen to $a / 16$ and 
the number of k points in the 2D reciprocal space is chosen as $4\times 4$. 

As for the time profile of the incident pulse, we use the following form:
\begin{eqnarray}
&&A^{\rm (i)}(z,t) = -\frac{1}{\omega} E_0 \, \mathrm{sin} \left[\omega \left(t-\frac{z}{c} -\frac{T}{2} \right) \right]  \nonumber \\ 
    && \times \mathrm{sin}^2 \left[ \frac{\pi }{T} \left(t-\frac{z}{c}\right) \right],\, (0<t-z/c<T),
\end{eqnarray}
where $E_0$ is the maximum amplitude of the electric field, $\omega$ is the average frequency, and $T$ is the pulse duration. 
We start the calculation at a time when the entire incident pulse locates outside the box area.
We will use the pulse duration of $T=18$ fs and the frequency in the range of $\hbar \omega =$ 1-- 9 eV. 
The time step is set to $\Delta t_{\rm e}=1.25\times 10^{-3}$ fs and $\Delta t_{\rm EM} = \Delta t_{\rm e}/100$.

We assume an adiabatic approximation for the exchange-correlation potential and 
use the local density approximation \cite{Perdew1981}.
We ignore the exchange-correlation effects for the vector potential in the present calculation, for simplicity.

\begin{figure}
    \includegraphics[keepaspectratio,width=\columnwidth]{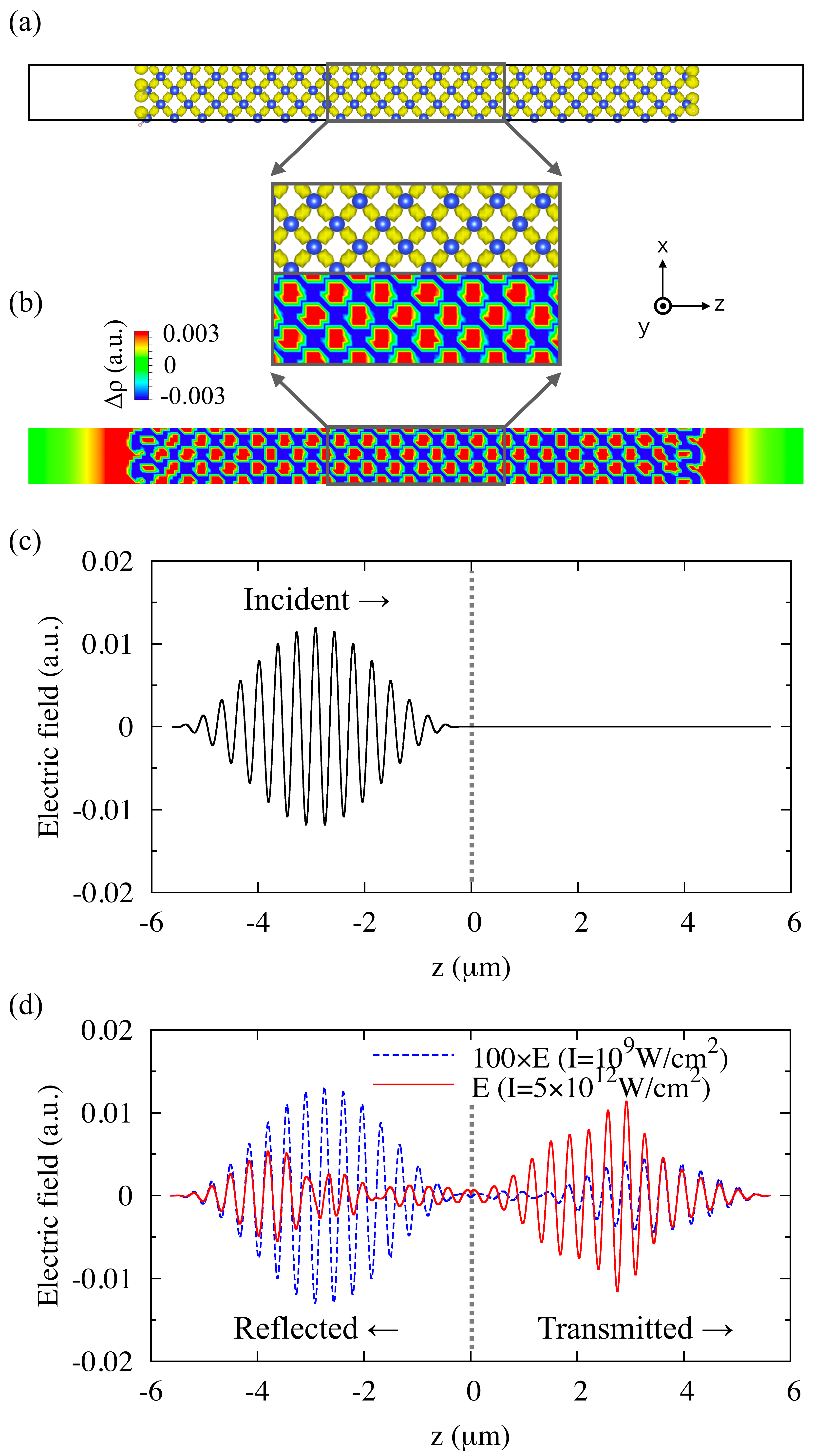}
    \caption{\label{fig:code} 
    Typical result of the microscopic Maxwell-TDDFT calculation with the $d=10a$ thin film.
    The average light frequency is $\hbar\omega=3.5$ eV. The intensity of the light pulse is $I=5\times10^{12}$ W/cm${}^2$.
    (a) Electron density in the ground state, visualized by VESTA\cite{Momma2011}. The enlarged view of the indicated area is also shown.
    (b) Change of the electron density from the ground state due to the pulse irradiation.
    (c) Incident light pulse and the target film ($z=0$).
    (d) Reflected and Transmitted wave (red solid line) after the pulse passed through the film ($t=18.75$ fs).
    For comparison, the results of weak pulse case ($I=10^{9}$ W/cm${}^2$) is also plotted as the blue dashed line. The latter is scaled up by a factor of 100.
    }
\end{figure}

In Fig.~\ref{fig:code}, we show a typical result of the microscopic Maxwell-TDDFT calculation. 
The thickness of the Si film is set to $d=10a \sim 5.4$nm and the frequency is $\hbar \omega = 3.5$ eV.
This frequency is above the optical gap energy of bulk silicon and a strong absorption of the light pulse is expected.
We show the results of two different maximum amplitudes, a sufficiently weak case in which a linear response theory
is applicable, and a strong case in which nonlinear light-matter interaction is significant.
Panel (a) shows the electron density in the ground state. 
High-density region is seen in the bond region.
The electric field of the incident pulse is shown in (c) for the case of the strong pulse.
The Si film is placed at $z=0$ position. Since the thickness of the Si film is much smaller than the 
wavelength of the incident pulse, the Si film is drawn as a dashed line without thickness in the figure.

In panel (d), the electric field after the interaction is shown for two cases. The incident pulse split into 
transmitted and reflected waves. Blue dotted curves indicate the results of weak pulse case, 
while red-solid curves show the results of the strong pulse case. The electric field of the weak pulse case is 
rescaled so that it can be directly compared with that of the strong pulse case.
The transmitted wave of the strong pulse case is slightly more intense than that of the weak pulse case.
On the other hand, the reflected wave of the strong pulse case is much smaller than that of the weak pulse
case. This fact indicates that the nonlinear light-matter interaction for the strong incident pulse manifests 
in the strong enhancement of the absorption and the strong reduction of the reflection by the Si film.

In Panel (b), the change of electron density from the ground state is shown for the case of strong pulse incidence at the time when the electric field is shown in the panel (d). 
The density change is shown after averaging over the $y$ direction.
We can see that the electron density of $\sigma$ bonds spreads out
and fill the spatial region outside the bonds. The absorbed energy is
spent to move the binding electrons and to weaken the bonds. 

\section{Low intensity case: Linear regime \label{sec:low}}

In this section, we investigate the interaction of a weak light pulse of the maximum intensity $I=10^{9}$ W/cm${}^2$ with the Si thin films of various thickness.
At this intensity, the linear response theory can be applied.
As explained in Secs.~\ref{sec:2D_macro} and \ref{sec:3D_macro}, there are two regimes in which we can introduce
macroscopic description utilizing constitutive relations: 2D macroscopic description 
for the thin limit and ordinary, 3D macroscopic description for the thick limit.
We expect the microscopic Maxwell-TDDFT scheme includes both limits and is applied to
films of intermediate thickness in which it is difficult to find a macroscopic description utilizing
a constitutive relation.

We consider there are two aspects that are important for the linear optical response of thin films. 
One is the change of the electronic structure in the thin films from that in the bulk.
The other is a multiple reflection inside the film. 
As is easily understood, the microscopic 
Maxwell-TDDFT scheme takes into account these two features in the formalism,
while two macroscopic descriptions, 2D developed in Sec.~\ref{sec:2D_macro} and 3D developed in Sec.~\ref{sec:3D_macro} are not.
In the 2D macroscopic description of Sec.~\ref{sec:2D_macro}, electronic structure of the thin films can be
taken into account through the 2D electric susceptibility of Eq.~(\ref{eq:2D_chi}). 
However, the effect of the multiple reflections is not since we take the zero-thickness limit. 
In the 3D macroscopic description of Sec.~\ref{sec:3D_macro}, the effect of the multiple reflections is taken into account, 
while the effect of the structure changes is not since the bulk electric susceptibility is used.

\subsection{2D electric susceptibility}

\begin{figure}
    \includegraphics[keepaspectratio,width=\columnwidth]{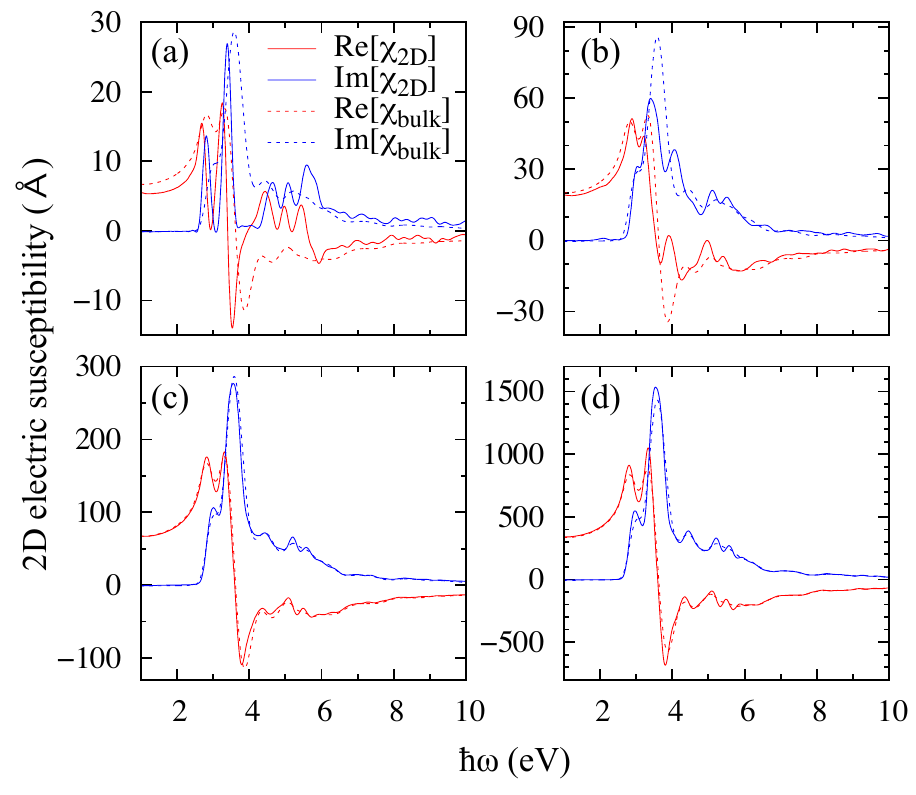}
    \caption{\label{fig:chi} 
    2D electric susceptibility as a function of the frequency $\hbar \omega$.
    The panels (a)--(d) correspond to the films of thickness $d=a$, $3a$, $10a$, and $50a$, respectively.
    The red (blue) solid lines show the real (imaginary) part of $\tilde \chi(\omega)$, and the dashed lines show  
    $\tilde \chi_{\rm bulk}(\omega) = \chi_{\rm bulk}(\omega)d$ .
    }
\end{figure}

We first investigate the influences of the changes of the electric structure of thin films from the bulk material
on the 2D electric susceptibility defined by Eq.~(\ref{eq:2D_chi}).
In Fig.~\ref{fig:chi}, we show the 2D electric susceptibilities $\tilde \chi(\omega)$ for the Si films of various thickness. 
Panels (a)--(d) correspond to films of thickness 
$d=a$, $3a$, $10a$, and $50a$, respectively.
For comparison, the electric susceptibility of the bulk silicon multiplied by the film thickness, 
$\tilde \chi_{\rm bulk}(\omega) = \chi_{\rm bulk}(\omega) d$, is also shown.
We note that this formula is justified in the thick limit as follows\cite{Li2018}:
If the polarization density $P({\bf r},\omega)$ is spatially homogeneous inside the film, 
the 2D polarization density $\tilde P(\omega)=\int dz P(\omega)$ is given by $P(\omega) d$ 
[see Eq.~(\ref{eq:jav})]. 
Then the 2D electric susceptibility is given by $\tilde \chi_{\rm bulk}(\omega) = \chi_{\rm bulk}(\omega) d$.

A comparison between the 2D and the 3D electric susceptibilities indicates the importance
of the structure change as the thickness decreases.
As seen from the figure, the difference is very small for sufficiently thick material, $d \ge 10a$.
As the thickness decreases, the difference becomes more and more significant.
At the thinnest case of $d = a$ in which the thickness is equal to the length of the single cubic cell, 
the difference is quite significant.
We can see that the quantum confinement effect sharpens the peaks in $\hbar \omega=$ 2--6 eV region.
We also find that the nearly zero plateau in the imaginary part appears around $\hbar \omega=$ 4 eV.

\subsection{Reflection, transmission, and absorption rates}

\begin{figure*}
    \includegraphics[keepaspectratio,width=\textwidth]{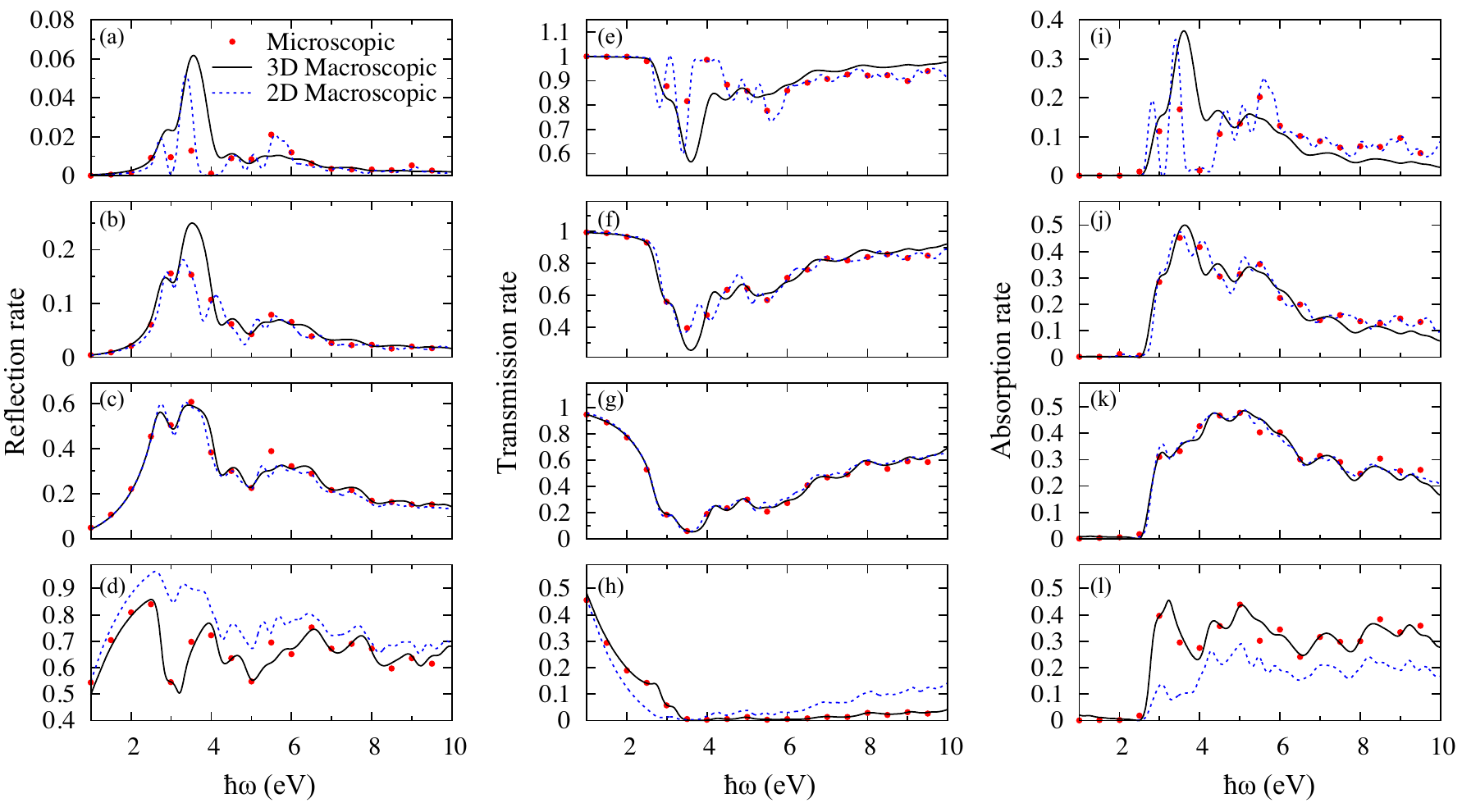} 
    \caption{\label{fig:abs_thickness}
    Reflection [(a)--(d)] , transmission [(e)--(h)], and absorption [(i)--(l)] rates of the films of thickness $d=a$, $3a$, $10a$, and $50a$, respectively, as a function of the frequency $\hbar \omega$.
    The red circles, black solid line, and blue dashed line correspond to the microscopic Maxwell-TDDFT, 3D macroscopic, and 2D macroscopic calculations, respectively.
    }
\end{figure*}

In Fig.~\ref{fig:abs_thickness}, we show the reflection, transmission, and absorption rates
as a function of the frequency for the weak light pulse irradiating on the Si films of various thickness.
We compare microscopic Maxwell-TDDFT calculations shown by red-circles with the
rates using 2D and 3D macroscopic descriptions explained in Sec.~\ref{sec:2D_macro} and Sec.~\ref{sec:3D_macro}, respectively.

In the microscopic Maxwell-TDDFT calculations, the rates are calculated using 
$e^{\rm (r)}_{\rm EM}(t)$ and $e^{\rm (t)}_{\rm EM}(t)$ defined by Eqs.~(\ref{r-Poynting}) and (\ref{t-Poynting}), respectively,
at the time $t=18.75$ fs, immediately after the pulse passed through the film.
We will later discuss the time dependence of the rates $e^{\rm (r,t)}_{\rm EM}(t)$ that indicate the delayed 
emission (Sec.~\ref{sec:emission}).

Here both effects of the changes of the electronic structure and the multiple reflections
appear in the comparison.
In the 2D macroscopic description, the electronic structure changes are incorporated
through the 2D electric susceptibility as shown in the previous subsection, while the multiple reflections are not.
In the 3D macroscopic description, while the multiple reflections are correctly taken into
account, the electronic structure changes are not.
We note that the rates of the microscopic Maxwell-TDDFT calculations are for light pulses which include finite frequency width, while those of the 2D and the 3D macroscopic descriptions are for monochromatic waves.

As seen from the figure, calculations using the 2D electric susceptibility coincide accurately 
with the microscopic Maxwell-TDDFT calculations for the films of thickness except for 
the thickest case, $d=50a$. This indicates that the description using the 2D electric susceptibility, 
Eq.~(\ref{eq:2D_chi}), is successful for films of thickness of 5 nm and less. 
In the $d=50a$ case, the absorption using the 2D electric susceptibility is very small compared 
with the other two calculations. The failure of the 2D description for this case indicates the
importance of multiple reflection for that thickness.

The description using the 3D electric susceptibility accurately coincides with the microscopic
Maxwell-TDDFT calculations for thicker films, as may be expected.
The difference between the three calculations is very small at $d=10a \sim 5$ nm.
As the thickness decreases, the difference becomes large, and the calculation using
the bulk electric susceptibility is poor for the thinnest case of $d=a \sim 0.5$ nm.
This is reasonable because the electronic structure of such thin film is very different from
the bulk one, as seen in the comparison of the electric susceptibilities.

\subsection{Spatial dependence of energy deposition}

\begin{figure}
    \includegraphics[keepaspectratio,width=\columnwidth]{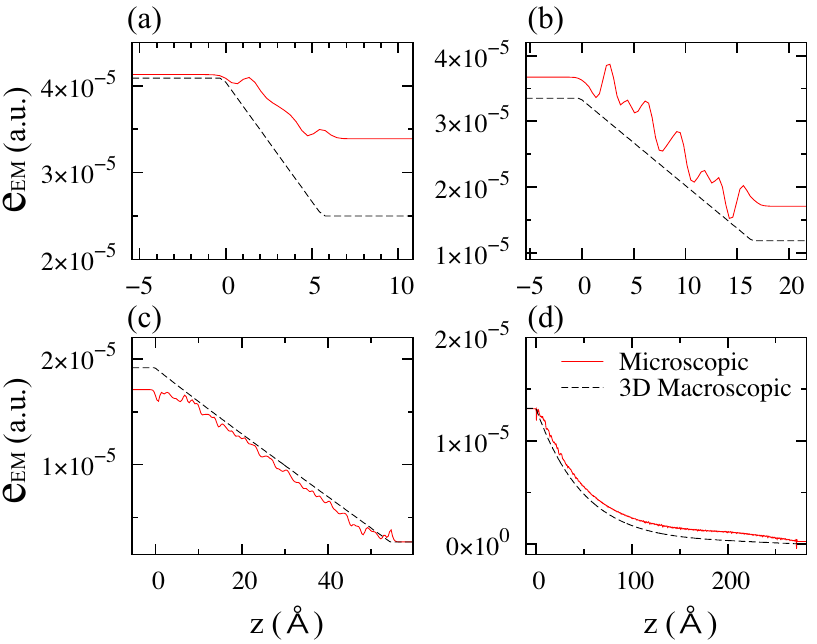}
    \caption{\label{fig:poy}
    Transmitted energy per area of the electromagnetic field [Eq.~(\ref{eq:e_Poynting})] as a function of $z$ at the time $t=18.75$ fs.   
    The panels (a)--(d) correspond to the films of thickness $d=a$, $3a$, $10a$, and $50a$, respectively.
    The red solid (black dashed) line corresponds to the microscopic Maxwell-TDDFT (3D macroscopic) description. 
    The light frequency is set to $\hbar\omega=3.5$ eV.  }
\end{figure}

\begin{figure}
    \includegraphics[keepaspectratio,width=\columnwidth]{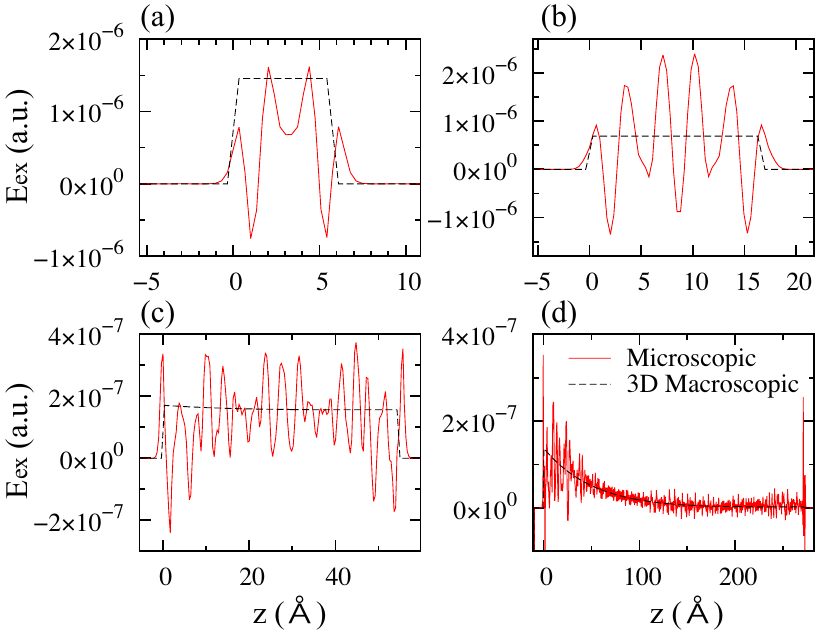}
    \caption{\label{fig:ex}
    Excitation energy [Eq.~(\ref{eq:excitation})] as a function of $z$.
    The conditions are the same as those of Fig.~\ref{fig:poy}.}
\end{figure}

To get a better understanding of the mechanism, we investigate the energy deposition, that is, 
the energy transfer from the light pulse to the electrons in the film.  
A comparison is made for two calculations, ordinary macroscopic description using  the 3D 
electric susceptibility and the microscopic Maxwell-TDDFT calculations. 
For the 2D macroscopic description, it is not possible to get such information, since we take the zero-thickness limit.

In the microscopic Maxwell-TDDFT calculation, the energy deposition is given as the derivative of the transmitted energy 
per area $e^{}_{\rm EM}(z,t)$ defined by Eq.~(\ref{eq:e_Poynting}) with respect to $z$:
\begin{equation}
    E_{\rm ex}(z,t) = - \frac{\partial}{\partial z} e^{}_{\rm EM}(z,t).
    \label{eq:excitation}
\end{equation}

Figure~\ref{fig:poy} and Figure~\ref{fig:ex} show the $z$ dependence of $e^{}_{\rm EM}(z,t)$ and $E_{\rm ex}(z,t)$ 
at the time $t=18.75$ fs at which the incident pulse ends, respectively. 
Results of  the microscopic Maxwell-TDDFT calculation show a rapid oscillation as a function of $z$ coordinate,
which reflects the atomic layers in the thin film, while the 3D macroscopic calculation is smooth.
Except for the thinnest case of $d=a$, the average behavior of the energy deposition of the microscopic
Maxwell-TDDFT calculation coincides with the energy deposition of the 3D macroscopic description.
This is consistent with the results of the rates shown in Fig.~\ref{fig:abs_thickness}.
For the thin films of $d=a$ and $d=3a$, the energy deposition is almost constant in space in the film. 
This indicates that the electromagnetic fields inside the medium are almost uniform in such very thin film. 
As the thickness increases, the energy deposition starts to show a decrease as $z$ increases.
This indicates that the light pulse is attenuated as it propagates through the thin film.
At the thickest case of $d = 50a$, the energy deposition shows almost exponential decrease. 

This finding explains the reason for the failure of the 2D macroscopic description for thick films.
The propagation effect becomes significant as the thickness increases, and the spatial dependence of the electric
field in the $z$ direction becomes sizable for the thickest case of $d=50a$. 
For such case, the 2D macroscopic description that ignores the spatial dependence cannot be validated.

\subsection{Delayed emission \label{sec:emission}}

\begin{figure}
    \includegraphics[keepaspectratio,width=\columnwidth]{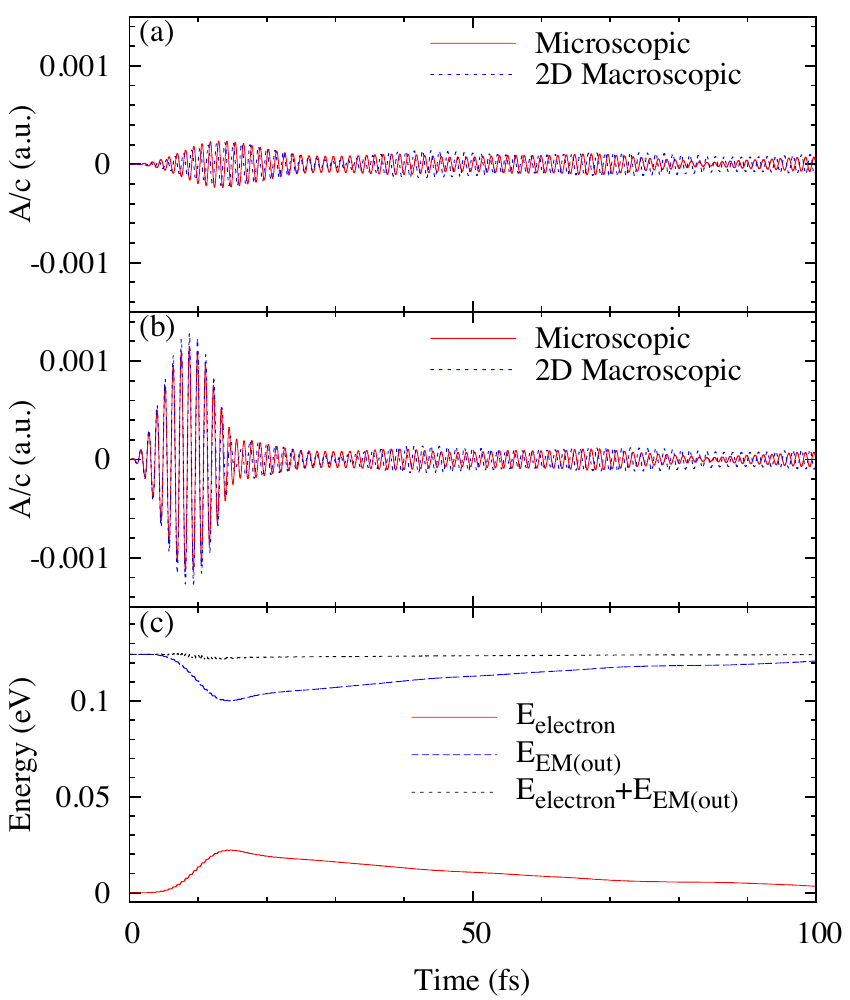}
    \caption{\label{fig:Ac_energy}
    Vector potential of the reflected [panel (a)] and transmitted [panel (b)] waves as a function of time for the $d=a$ thin film.
    The average frequency is $\hbar \omega=3.5$ eV.
    In the panel (a) and (b), the red solid (blue dotted) line corresponds to the microscopic Maxwell-TDDFT (2D macroscopic) result. 
    Panel (c): the electronic excitation energy of the thin film (red solid line), the energy of the electromagnetic field outside the calculation box area (blue dashed line), and the sum of the two (black dotted line) are shown.
    }
\end{figure}

\begin{figure}
    \includegraphics[keepaspectratio,width=\columnwidth]{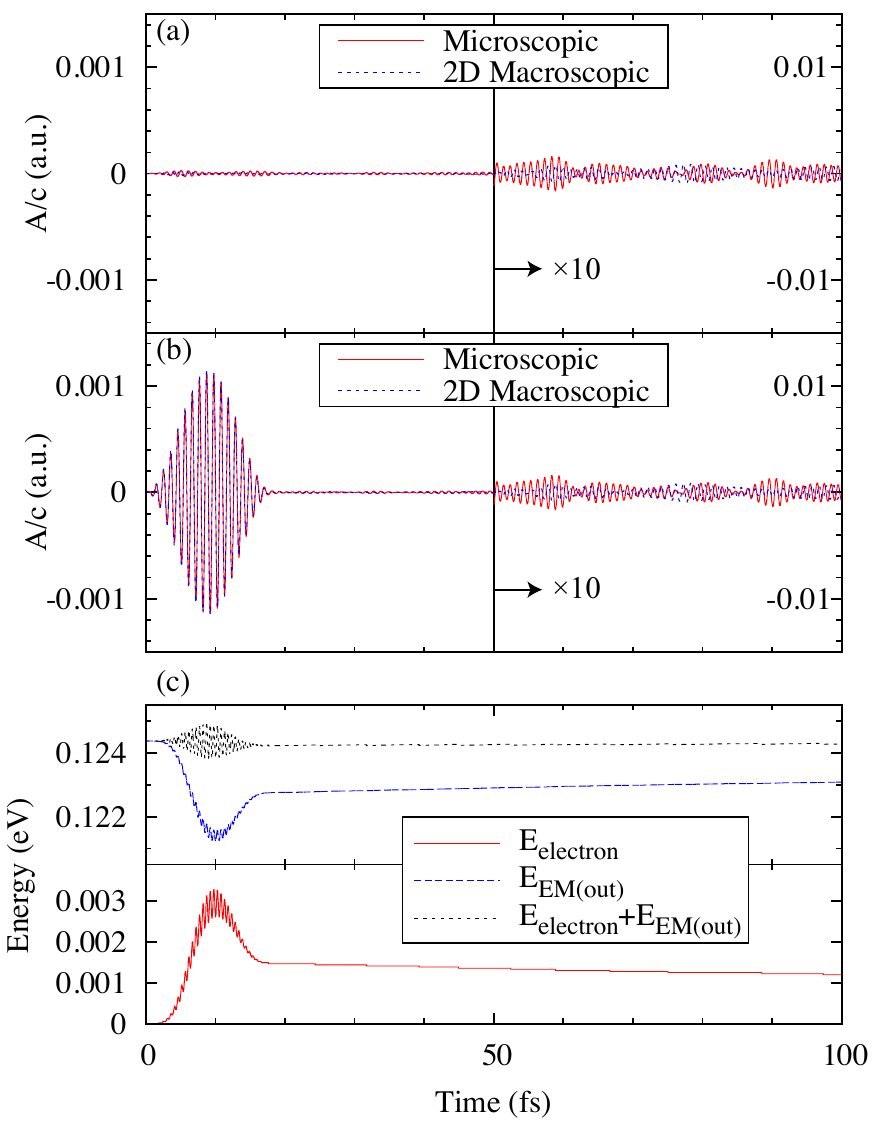}
    \caption{\label{fig:Ac_energy_4eV}
    The same as Fig.~\ref{fig:Ac_energy} but $\hbar \omega=4$ eV.
    In the panel (a) and (b), the vector potentials are scaled up by a factor of 10 after the time $t=50$ fs.
    }
\end{figure}

In the results shown up to here, we used the energy transmission function, $e_{\rm EM}(z,t)$ of Eq.~(\ref{eq:e_Poynting}) 
at the time $t=18.75$ fs.
This means that we calculate the reflection, and the transmission rates shown in Fig.~\ref{fig:abs_thickness} 
from the electromagnetic fields emitted outside the thin film immediately after the end of the irradiation of the
incident pulse. 

In the microscopic Maxwell-TDDFT calculation, the energy once stored in the electrons in the medium is re-emitted
as light electromagnetic fields, which we call the delayed emission and will be discussed in this subsection.
In panels (a) and (b) of Fig.~\ref{fig:Ac_energy} and Fig.~\ref{fig:Ac_energy_4eV}, we show a vector potential of the reflected and 
transmitted waves, $A^{\rm (r)}(t)$ and $A^{\rm (t)}(t)$, respectively, as a function of time for a long time span.
The red solid line and the blue dotted line indicate the microscopic Maxwell-TDDFT calculation and the 2D macroscopic description [Eqs.~(\ref{eq:2DAmacro}) and (\ref{eq:J_sigmaE})], respectively.
The thickness of the thin film is $d=a$, and the average frequency is $\hbar\omega = 3.5$ eV and $\hbar\omega = 4$ eV 
in Fig.~\ref{fig:Ac_energy} and Fig.~\ref{fig:Ac_energy_4eV}, respectively.
We see that the light emission continues a long time after the incident light pulse passes through the thin film.
The delayed emission appears both transmitted and reflected waves, and the pulse shapes are very similar to each other. 
This is understood from the 2D macroscopic description of Sec.~\ref{sec:2D_macro}.
The continuity equation of Eq.~(\ref{eq:2Dcont}), $A^{\rm (i)}(t) + A^{\rm (r)}(t) = A^{\rm (t)}(t)$, indicates that
the transmitted and the reflected waves should be the same after the incident pulsed field ends.
The 2D macroscopic result has roughly the same behavior with the microscopic Maxwell-TDDFT result.

In panel (c) of Fig.~\ref{fig:Ac_energy} and Fig.~\ref{fig:Ac_energy_4eV}, we show a decomposition of the
energy of the system. 
The red-solid lines indicate the electronic excitation energy in the thin film.
The blue-dashed line is the energy of the electromagnetic fields outside the computational box area, and the black dotted
line is the sum of the two. The summed energy is not completely constant value, because there is an
energy of the light electromagnetic fields in the thin film.
As seen from the figure, the electronic excitation energy changes into the electromagnetic field in the
delayed emission process.
We find that the electronic excitation energy, or equivalently, the absorption rate derived by $e_{\rm EM}(z,t)$ eventually goes to zero in the infinite future.
The electronic excitation as well as the delayed emission is much larger in the case of $\hbar\omega=3.5$ eV,
shown in Fig.~\ref{fig:Ac_energy}, than those in the case of $\hbar\omega=4.0$ eV shown in Fig.~\ref{fig:Ac_energy_4eV}.
This can be understood from the 2D electric susceptibility shown in Fig.~\ref{fig:chi}(a). The imaginary part of the
susceptibility shows a sharp peak at 3.5 eV, while it is very small at 4.0 eV for the  film of $d=a$ thickness.

The emission seen in Fig.~\ref{fig:Ac_energy} is related to the very sharp structure at around 3.5 eV 
in Fig.~\ref{fig:chi}(a).  This indicates the following mechanism for the delayed emission:
After irradiating the thin film with a pulse of the average frequency of 3.5 eV, the thin film is
highly excited due to the resonant excitation mechanism. Then, through a coupling with the
emission channel of the electromagnetic fields, the delayed emission takes place.
In fact, taking the Fourier transform of the emitted pulse, it produces a peak at the same
energy and the same width as that seen in the 2D electric susceptibility. Therefore, the time scale of
the emission is related to the width of absorption peak through the uncertainty principle.

We should, however, note that whether such delayed emission can be observed experimentally depends on the mechanisms of dephasing and decays that take place through the electron-phonon coupling and electron-electron scattering are not taken into account sufficiently in the present scheme.

\section{High intensity light pulse: Nonlinear regime \label{sec:high}}

\begin{figure}
    \includegraphics[keepaspectratio,width=\columnwidth]{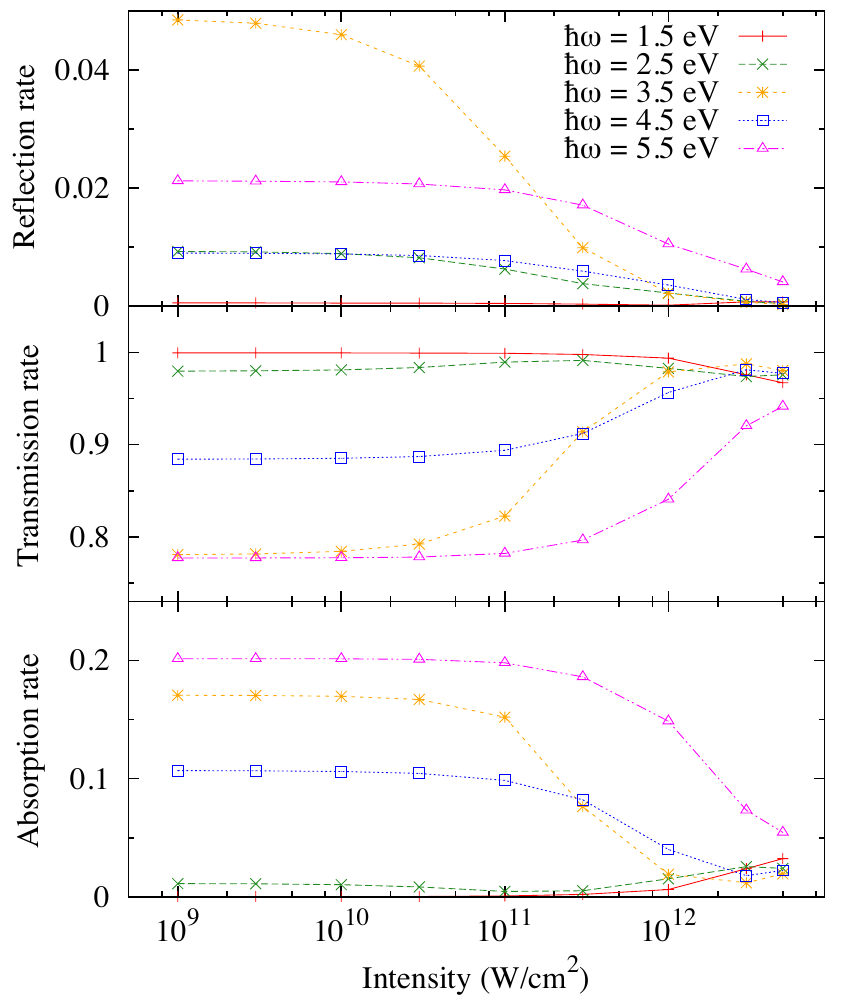} 
    \caption{\label{fig:rta_intensity}
    Reflection, transmission, and absorption rates of the film of thickness $d=a$ as a function of the light intensity, for several average frequencies.
     }
\end{figure}

\begin{figure*}
    \includegraphics[keepaspectratio,width=\textwidth]{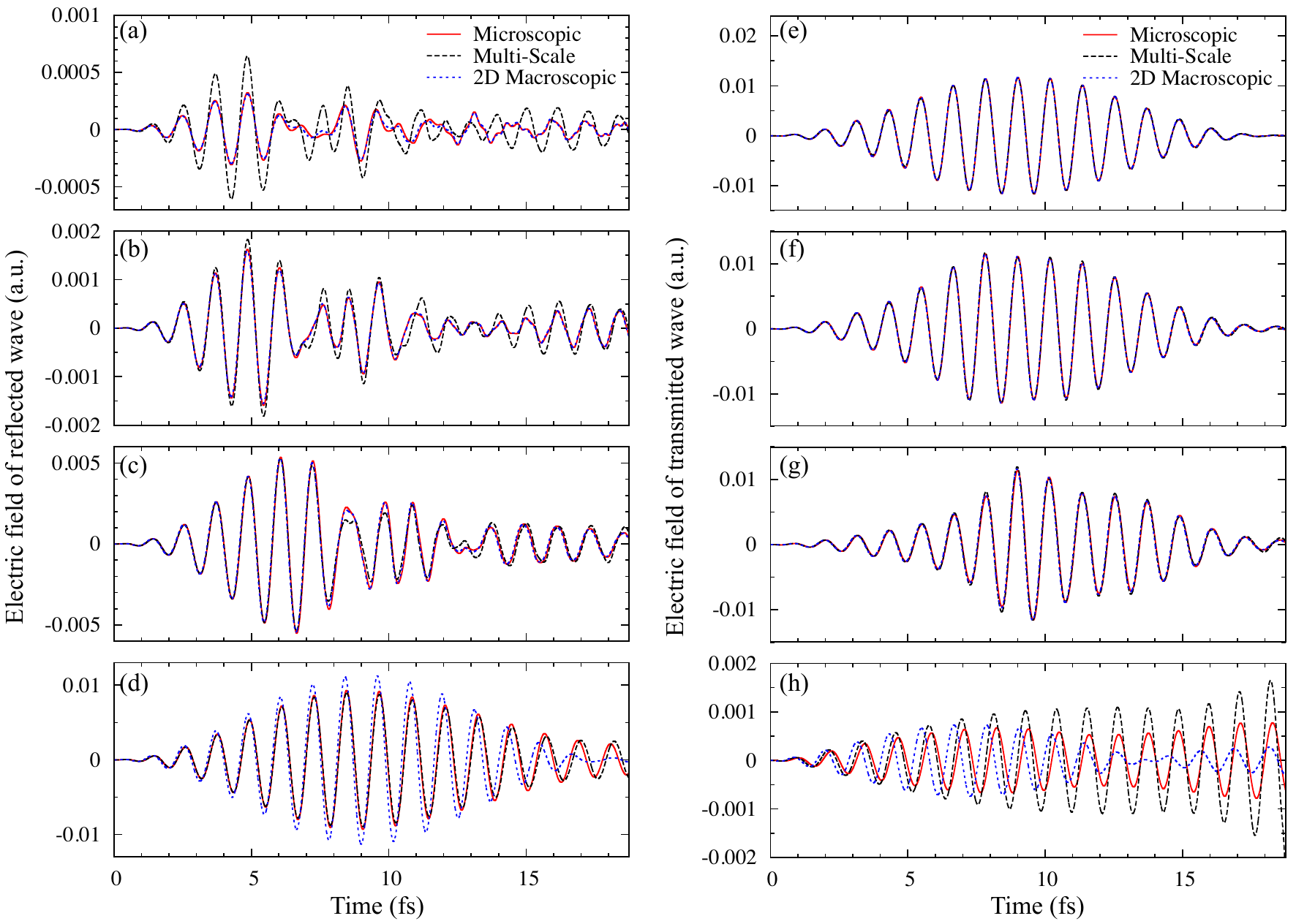}
    \caption{\label{fig:5d12_ref_t}
    Reflected wave [(a)--(d)] and transmitted wave [(e)--(h)] as a function of the time. 
    The panels (a)--(d) [(e)--(h)] correspond to the films of thickness $d=a$, $3a$, $10a$, and $50a$, respectively.
    The red solid line, black dashed line, and blue dotted line corresponds to the microscopic Maxwell-TDDFT, 3D macroscopic (multiscale Maxwell-TDDFT), and 2D macroscopic calculations, respectively.
    The average frequency is $\hbar \omega=3.5$ eV and the light intensity is $I=5\times10^{12}$ W/cm${}^2$.
    }
\end{figure*}

In this section, we consider nonlinear interaction in thin materials when an intense and ultrashort
light pulse irradiates normally on the material. 
We first show a systematic view for the nonlinear interaction for the very thin film of $d=a$ thickness 
using the microscopic Maxwell-TDDFT scheme.
In Fig.~\ref{fig:rta_intensity}, reflection, transmission, and absorption rates are shown for pulses of various maximum
intensities and for several average frequencies.
The interaction is in linear regime below the intensity of $1.0 \times 10^{10}$ W/cm$^2$, 
where the rates do not depend on the intensity. 
Above $1.0 \times 10^{10}$ W/cm$^2$, the rates changes substantially: the reflection and the absorption rates 
decrease and the transmission rate increases. 
This indicates that the intense light pulse tends to transmit through the thin film for light pulse of intensities more than $1.0 \times 10^{10}$ W/cm$^2$.
Looking in more detail, there is a difference in the intensity at which the nonlinearity starts to appear:
For a light pulse of the average frequency 3.5 eV where a strong absorption is seen in the 2D electric susceptibility, the rates start to deviate at the lowest intensity around $1.0 \times 10^{10}$ W/cm$^2$.
This suggests the saturable absorption as the origin of the nonlinearity: a sizable transition in a specific
orbitals/k-points causes the saturation of the transition due to the lack of electrons in the occupied orbitals
and/or the blocking of the unoccupied orbitals.

We next investigate how the macroscopic descriptions work (Fig.~\ref{fig:5d12_ref_t}).
We compare three calculations, the microscopic Maxwell-TDDFT, the 2D macroscopic calculation solving
Eqs.~(\ref{eq:TDKS2D}) and (\ref{eq:2DAmacro}) simultaneously, and the multiscale Maxwell-TDDFT 
calculations solving Eqs.~(\ref{eq:Aeqmultiscale}) and (\ref{eq:TDKS-3D}) simultaneously,
for the films of various thickness.

We first discuss the reflected wave [Fig.~\ref{fig:5d12_ref_t}(a)--(d)].
At the thinnest case of $d=a$, the amplitude of the reflected
wave is very small. The microscopic Maxwell-TDDFT calculation coincides accurately with the
2D macroscopic calculation. This indicates that the 2D macroscopic calculation is reliable for
this thickness. On the other hand, multiscale Maxwell-TDDFT calculation fails to reproduce the
microscopic Maxwell-TDDFT calculation. 
This indicates the importance to treat the electronic
structure of a very thin material, as discussed in the linear response calculation.

As the thickness increases, three calculations more or less coincide with each other. 
At the thick limit of $d=50a$, calculations of multiscale Maxwell-TDDFT and
microscopic Maxwell-TDDFT show good coincidence with each other. 
However, the 2D macroscopic calculation fails. Therefore, the 2D macroscopic approach fails for
films of a thickness more than 5 nm.
These results are very similar to the case of the weak field: The microscopic Maxwell-TDDFT
scheme successfully describes the nonlinear interaction irrespective of the thickness of the film.
The 2D macroscopic description is successful for films of thickness 5 nm or less.
The multiscale Maxwell-TDDFT scheme, on the other hand, successful
for films of 5 nm or more.

For the transmitted wave [Fig.~\ref{fig:5d12_ref_t}(e)--(h)], three calculations coincide with each other up to $d=10a$ case.
In the thickest case shown in Fig.~\ref{fig:5d12_ref_t}(h), the transmitted wave is rather weak and three calculations look different.
The microscopic and multiscale calculations look similar at the beginning.
However, the difference gradually increases.

\section{Conclusion \label{sec:conclusion}}

We have developed a comprehensive and computationally feasible approach for  the  interaction of
an ultrashort light pulse and thin materials of various thickness.
Our description is based on first-principles time-dependent density functional theory.
In the most comprehensive description which we call microscopic Maxwell-TDDFT scheme, we couple the time-dependent Kohn-Sham equation,
the basic equation of the time-dependent density functional theory, and the classical
Maxwell equations for light electromagnetic fields using a common spatial grid. 
We also develop macroscopic descriptions for two limiting cases: two-dimensional macroscopic
electromagnetism for extremely thin materials and three-dimensional macroscopic
electromagnetism for sufficiently thick materials. 
Our schemes can be applied even when the incident field is very strong so that the light-matter interaction is highly nonlinear.

We find that the microscopic Maxwell-TDDFT scheme can describe the interaction
of light pulses of various intensities with thin films of various thickness. 
It incorporates effects of electronic structure changes of the thin film from the bulk material in the first
principles level, multiple reflection effects at both surfaces of the thin film, and the nonlinear light-matter interaction for strong incident pulses.

We find the two-dimensional macroscopic description works well for films of thickness 5 nm or less, while the three-dimensional macroscopic description works well for films of thickness 5 nm or more.

\begin{acknowledgements}
This research was supported by JST-CREST under grant number JP-MJCR16N5, and by MEXT as a priority issue theme 7 to be tackled by using Post-K Computer, and by JSPS KAKENHI Grant Numbers 15H03674. Calculations are carried out at Oakforest-PACS at JCAHPC under the support by Multidisciplinary Cooperative Research Program in CCS, University of Tsukuba.
\end{acknowledgements}

%


\end{document}